\documentclass[%
showpacs,
 amsmath,amssymb,
 aps,
 prl,
 lengthcheck,%
]{revtex4}

\usepackage{graphicx}
\graphicspath{{.},{./figs/}}
\usepackage{dcolumn}
\usepackage{bm}
\usepackage{hyperref}

\usepackage{color}

\newcommand{\mylab}[1]{\label{#1}}
\begin{document}

\title{Dynamical model for the formation of patterned deposits at
  receding contact lines}
\author{Lubor Frastia}
 \email{l.frastia@lboro.ac.uk}
\author{Andrew J.~Archer}
 \email{a.j.archer@lboro.ac.uk}
\author{Uwe Thiele}
  \email{u.thiele@lboro.ac.uk}
\affiliation{Department of Mathematical Sciences, Loughborough University, Loughborough, Leicestershire, LE11 3TU, UK}%

\date{\today}

\begin{abstract}
  We describe the formation of deposition patterns that are observed
  in many different experiments where a three-phase contact line of a
  volatile nanoparticle suspension or polymer solution recedes. A
  dynamical model based on a long-wave approximation predicts the
  deposition of irregular and regular line patterns due to
  self-organised pinning-depinning cycles corresponding to a
  stick-slip motion of the contact line. We analyze how the line
  pattern properties depend on the evaporation rate and solute
  concentration.
 \end{abstract}


\maketitle

%
%
The last decade has seen huge growth in interest in phenomena that
accompany evaporative and convective dewetting of suspensions and
solutions. Well known are the detailed studies of the coffee stain
effect \cite{Deeg97,Deeg00} that analyse the deposition and resulting
structures left behind by a receding three-phase contact line of an
evaporating drop of suspension upon a solid substrate.  In particular,
Ref.~\cite{Deeg00} describes a large range of different deposition
patterns including cellular and lamellar structures, single and
multiple rings, and Sierpinski gaskets. Other observed structures
include crack \cite{Dufr03} and chevron \cite{Bert10} patterns.
Recently it has been shown that evaporating polymer solutions
\cite{YaSh05,Xu06,HXL07} and (nano)particle suspensions
\cite{RDLL06,XXL07,BDG10} may be used to fabricate strikingly regular
stripe patterns, where the deposited stripes are parallel to the
receding contact line and have typical distances ranging from
10-100$\mu$m.  The goal is to use this effect as a non-lithographic
technique for covering large areas with regular arrays of small-scale
structures, such as, e.g., concentric gold rings with potential uses
as resonators in advanced optical communications systems
\cite{HXL06}. The deposited patterns from more complex fluids, such as
polymer mixtures \cite{Byun08} and DNA solutions \cite{MZZC08}, are
also investigated.
The occurrence of regular stripe patterns is a somewhat generic
phenomenon, that is not only observed for different combinations of
substances but also in a variety of experimental setups that allow for
slow evaporation. Examples include the meniscus technique in a
sphere-on-flat geometry \cite{XXL07,HXL07}, a controlled continuous
supply of liquid between two sliding plates to maintain a
meniscus-like surface \cite{YaSh05} and dewetting forced by a pressure
gradient \cite{BDG10}. Interestingly, besides the stripes parallel to
the receding contact line, a variety of other patterns are observed,
including regular orthogonal stripes \cite{XXL07}, superpositions of
orthogonal and parallel stripes \cite{YaSh05}, regular arrays of drops
\cite{KGMS99,YaSh05} and irregularly branched structures
\cite{KGMS99,Paul08}. This behaviour is highly sensitive to the
particular experimental setup and parameters.

%
Despite the extensive number and variety of experiments, an
explanation of the formation of the regular patterns has been rather
elusive. Although most studies agree that the patterns result from a
stick-slip motion of the contact line caused by pinning/depinning
events \cite{Deeg00,HXL06,Xu06,BFA09} no dynamical model of the
periodic deposition process exists.  Most models assume a permanently
pinned contact line (see \cite{OKD09,Witt09} and references therein)
and are therefore only able to describe the formation of a single line
deposit. A non-isothermal Navier-Stokes simulation shows depinning
from such a single line but no periodic deposits \cite{BFA09}. The
model of Ref.~\cite{WCM03} describes drop arrays formed via directed
dewetting of the solvent that are subsequently dried.

%
In this Letter we discuss a generic close-to-equilibrium model for the
evaporative and convective receding of a three-phase contact line of a
solution or suspension on a solid substrate.  We show that solely
having a viscosity that diverges at a critical solute concentration is
sufficient to trigger a self-organised periodic pinning-depinning
process that results in the deposition of regular line patterns.  The
model can easily be extended to incorporate other processes and in the
future may be employed to assess their influence on the basic
mechanism that we describe here.

\begin{figure}
\includegraphics[width=0.8\hsize]{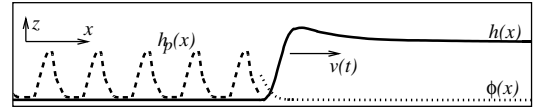}
\caption{Sketch of a liquid front that recedes due to evaporation and
  convection with a varying velocity $v(t)$. The deposition process is
  characterized by the film thickness profile $h(x,t)$, the
  concentration profile $\phi(x,t)$ (in the bulk film), and the
  particle layer thickness $h_p(x,t)=\phi(x,t) h(x,t)$ (outside the
  bulk film).}
\mylab{fig:sketch}
\end{figure}

We consider a thin film of an evaporating partially wetting
nanoparticle suspension (or polymer solution) in contact with its
vapour on a flat solid substrate (see Fig.~\ref{fig:sketch}).
Assuming that all surface slopes are small, one may employ a long-wave
approximation \cite{ODB97} and derive two coupled evolution equations
for the film thickness profile $h(x,t)$ and the vertically averaged
solute concentration field $\phi(x,t)$:
\begin{eqnarray}
\partial_th & =& \partial_x\left[Q(h,\phi)\partial_x p(h)\right]
 - \frac{\beta}{\rho}(p(h) - \mu\rho),
\mylab{e:tfeqh}\\
\partial_t(\phi h) &=& \partial_x\left[\phi Q(h,\phi)\partial_xp(h)\right]
 + \partial_x\left[D(\phi)h\partial_x\phi\right]. \mylab{e:tfeqhp}
\end{eqnarray}
The mobility $Q(h,\phi)=h^3/3\eta(\phi)$ models Poiseuille flow and
incorporates the dynamic viscosity $\eta(\phi)$ that exhibits a
strong non-linear dependence on the local solute concentration.
We employ the Krieger-Dougherty law \cite{Lars98,Quem77}
\begin{equation}
\eta(\phi)= \eta_0(1-\phi)^{-\nu},
\mylab{e:visc}
\end{equation}
where $\eta_0$ is the viscosity of the pure solvent. We have scaled
$\phi$ by the concentration at random close packing ($\phi_c=0.63$) so
that $\eta$ diverges when $\phi \to 1$.  The precise value of the
exponent $\nu = [\eta]\phi_c$ depends on the type of suspension. For
non-interacting particles (i.e.\ particles that have no net attractive
forces between them and only have excluded volume interactions),
values for $\nu$ between 1.4 and 3 are discussed, depending on the
shape of the particles \cite{Lars98}. For spherical particles the
factor $[\eta]=2.5$, giving $\nu = 1.575$. Other thin film models use
$\nu=2$ \cite{CBH08,WCM03}.  For interacting solute particles, values
for $\nu$ as low as 0.13 are reported \cite{Trap01}. Depending on the
particular system, the transition at $\phi_c$ is either referred to as
jamming or gelation \cite{Trap01}. Here we fix $\nu=1.575$, although
we have found that the effects we describe below are even stronger for
smaller $\nu$.

The first term on the right hand side of Eq.~(\ref{e:tfeqh})
(conserved part) corresponds to convective transport of the liquid
whereas the second term (non-conserved part) models evaporation. The
convective flow is driven by the gradient of the pressure
\begin{equation}
 p(h)= - \gamma\partial_{xx}h -\varPi(h),
\mylab{e:press}
\end{equation}
where the first term is the Laplace pressure ($\gamma$ is the surface
tension) and the second is the disjoining pressure $\varPi(h) =
2S^{LW}d_0^2/h^3 + S^{P}\exp[-(h-d_0)/l_0]/l_0$ that models a
partially wetting fluid \cite{deGe85,Shar93}. Here, $l_0$ is the
Debye length, $d_0$ is a molecular interaction length, $S^{LW} =
-A/12\pi d_0^2$ and $S^{P}<0$ are the apolar and polar spreading
coefficient, respectively, and $A<0$ is the Hamaker constant.
To derive the second term in Eq.~(\ref{e:tfeqh}) we assume the system
is close to equilibrium and near to saturation and so evaporation is
slow.  In this limit evaporation with a rate $\beta$ is driven by the
difference of the scaled pressure $p/\rho$ and the chemical potential
of the ambient vapour $\mu$ \cite{LGP02,Pism04}. Latent heat effects
may be neglected, and the density $\rho$ is assumed to be equal for
particles and solvent.
The first and second terms on the right hand side of
Eq.~(\ref{e:tfeqhp}) model convective and diffusive transport of the
particles, respectively.  Note that the diffusion coefficient depends
on concentration and we employ the Einstein-Stokes relation $D(\phi) =
k_BT/6\pi r_0\eta(\phi)$, where $k_B$ is the Boltzmann constant, $T$
the temperature, and $r_0$ the particle radius.

%
%
Models related to Eqs.~(\ref{e:tfeqh}) and (\ref{e:tfeqhp}) are used
in studies of particle-laden film flow \cite{CBH08} (without
evaporation or wettability effects) and dewetting of suspensions of
surface active particles \cite{WCM03} (different wettability
regime). In the limit $\phi \to 0$, our theory reduces to that used in
\cite{LGP02} to study the fingering instability of an evaporative
front of a pure liquid.  We choose our scaling and some of the
parameters to be the same as in \cite{LGP02}: The dimensionless
chemical potential $M=\rho\mu /|\tilde{S}^{P}|=-0.003$ and the
diffusion number $D_0=3k_BT/r_0[6\pi A^2|\tilde{S}^P|]^{1/3}=0.0003$,
where $\tilde{S}^P=S^P\exp(d_0/l_0)/l_0$.
Our main control parameters are the initial mean (dimensionless)
concentration $\phi_0$ and the evaporation number
$\varOmega_0=18\pi\beta\eta_0\gamma/\rho[6\pi A^2|\tilde{S}^P|]^{1/3}$
which represents the ratio of the time scales for convection and
evaporation of a film without solute.
%
%
We solve the nondimensional model by discretizing over a spatial
domain of finite length $L$. Deposited patterns are obtained by direct
time simulations using a variable-step variable-order backward
difference scheme starting from an initially step-like front that
becomes smooth in the early evolution \cite{FAT10_numdet}.

\begin{figure}[t]
\includegraphics[width=0.9\hsize]{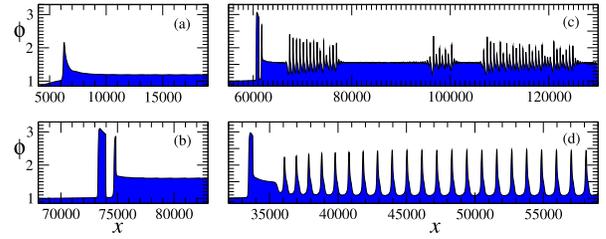}
\caption{(color online) Typical deposit profiles: (a) a single line,
  (b) a finite sequence of lines, (c) an intermittent line pattern,
  and (d) a regular line pattern.  The parameters are $\phi_0=0.41$
  and from (a) to (d): $\varOmega_0 = (14.7, 0.1, 0.147, 0.464)
  \times10^{-6}$.}
  \mylab{f:dewetting}
\end{figure}
In the evaporative dewetting process one encounters different types of
receding fronts. In the case without solute \cite{LGP02} one may
distinguish the limiting cases of (i) \textit{convection-dominated}
and (ii) \textit{evaporation-dominated dewetting} for small and large
values of $\varOmega_0$, respectively.  In case (i) the front recedes
rapidly and convective motion maintains a capillary ridge despite
evaporation.  In case (ii) convection is much slower than evaporation,
the front recedes slowly and there is no capillary ridge.
In the presence of a solute the situation is more complex and
stationary receding front shapes are not found for some parameter
values.  In general, on starting with a uniform concentration
$\phi(x,0) = \phi_0$, as the front recedes, it deposits part of the
solute in a smooth layer.  Evaporation in the contact line region
increases the local concentration $\phi$ and consequently also the
viscosity. When $\phi \to 1$, the convective motion in the contact
region stops completely (the suspension becomes locally jammed) due to
the strong non-linearity in Eq.\ (\ref{e:visc}).
One may say that the initial convection-dominated stage changes into
an evaporation-dominated regime as the local evaporation number
$\varOmega=(\eta(\phi)/\eta_0)\varOmega_0$ becomes large. At this
stage the front seems pinned. However, it is actually still moving
extremely slowly, solely by evaporation, and deposits a first line of
solute.  During this, the local concentration in the contact region
decreases, i.e., $\varOmega$ decreases, until the front depins and
convective motion resumes.  Subsequently, after the initial line is
deposited, various scenarios are possible: (a) a deposit of uniform
thickness (i.e.\ only a single line is deposited), (b) deposition of a
finite number of lines followed by a layer of constant thickness, (c)
intermittent line pattern, and (d) regular line pattern. In the narrow
region of parameter space where the `limiting' case (c) occurs, we
find that this behaviour is normally very sensitive to computational
details (which is not the case for (d)), so we believe that the
intermittent line patterns represent a `chaotic deposition'. Typical
profiles are displayed in Fig.~\ref{f:dewetting} and the corresponding
parameter ranges are marked in the phase diagram
Fig.~\ref{f:phasediag}.

\begin{figure}
\includegraphics[width=0.9\hsize]{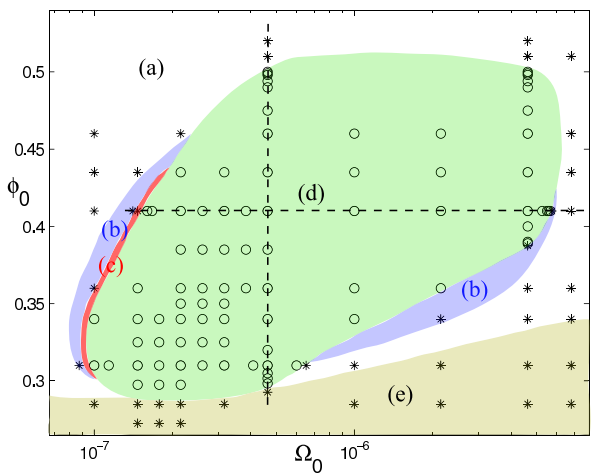}
\caption{(color online) Morphological phase diagram of deposition
  patterns, in the plane spanned by evaporation number $\varOmega_0$
  and bulk concentration $\phi_0$. Symbols denote performed
  simulations.  In the central region we find (d) regular line
  patterns ($\circ$) and outside of this region ($\ast$) we observe:
  (a) single lines, (b) multiple lines, (c) intermittent patterns, and
  (e) no lines.  For typical deposit profiles for (a)-(d) see
  Fig.~\ref{f:dewetting}. Results from along the dashed lines are
  analysed at Figs.~\ref{f:vertcut} and \ref{f:horizcut}.}
\mylab{f:phasediag}
\end{figure}
Ignoring initial transients, we distinguish two main types of
deposits: those of uniform thickness (regions (a), (b) and (e) in
Fig.~\ref{f:phasediag}) and periodic line patterns (region (d)).  In
the latter one observes (as in the experiments
\cite{HXL06,Xu06,BDG10}) a regular stick-slip motion of the contact
line, since the typical time scales for convection-dominated and
evaporation-dominated front motion may differ by orders of magnitude.

Structure formation results from a subtle interplay between
convection, evaporation, and diffusion. The basic mechanism of line
deposition described above stems from a balance of convective and
evaporative motion.  Diffusion does not change that picture as long as
its time scale is not much shorter than the convective and evaporative
time scales. Increasing $D_0$ merely shrinks the region (d) in
Fig.~\ref{f:phasediag}. However, in the unlikely situation that
diffusion is so fast that nanoparticles diffuse away ahead of the
receding contact line (e.g.\ when $D_0=0.3$), then line deposition is
suppressed. We do not consider this case here.

\begin{figure}[t]
\includegraphics[width=0.488\hsize]{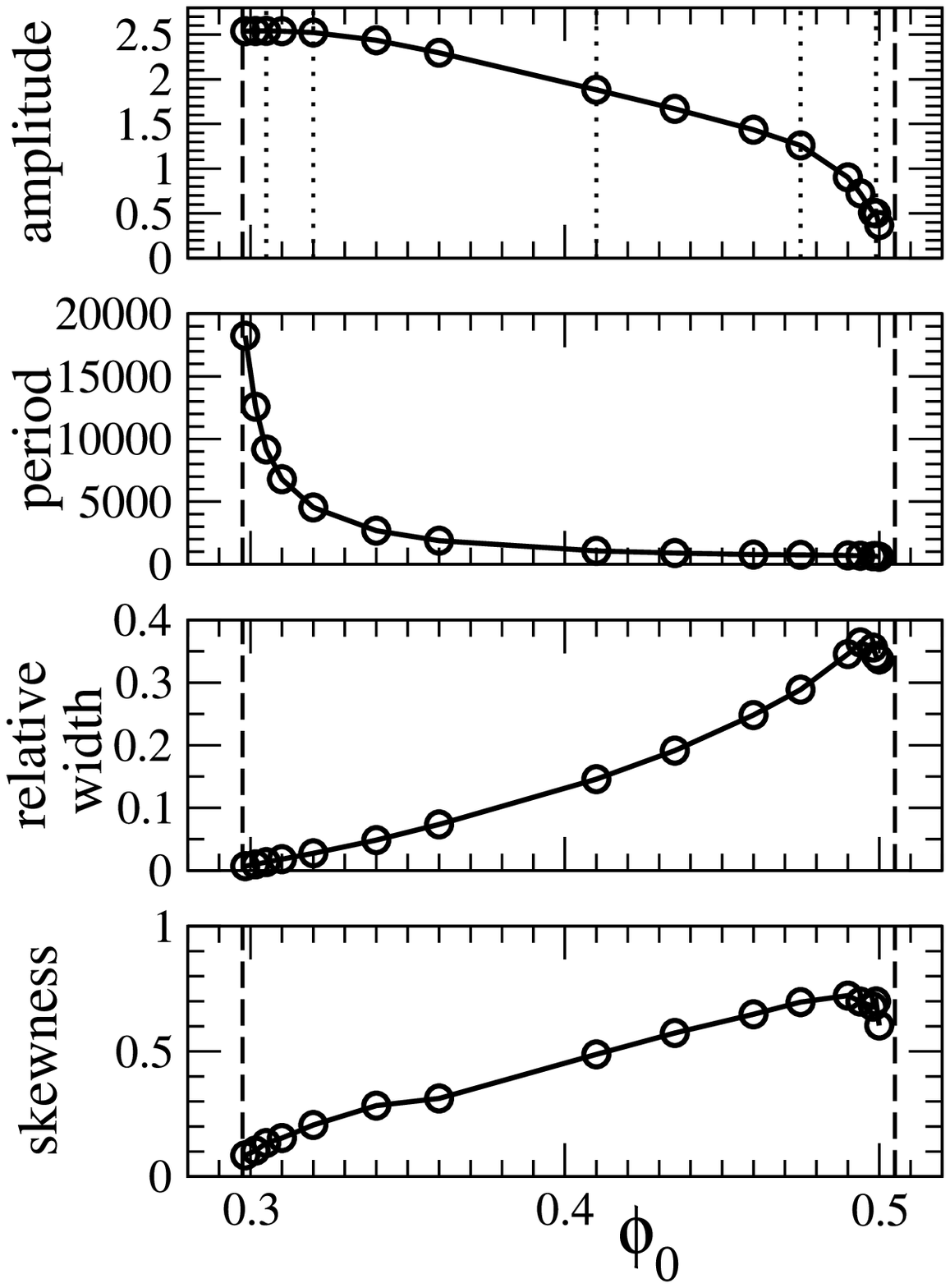}\hfill
\includegraphics[width=0.482\hsize]{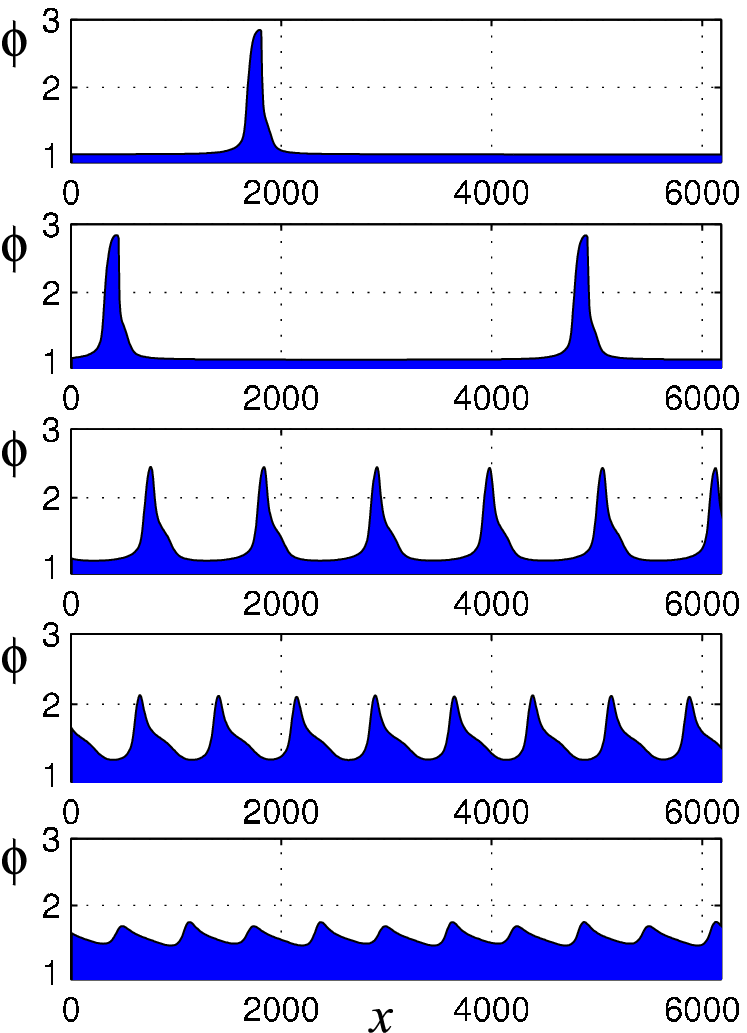}
\caption{(color online) Left: dependence of the regular line pattern
  properties on $\phi_0$ in the range bounded by the vertical dashed lines
  that corresponds to region (d) of Fig.~\ref{f:phasediag}.
  $\varOmega_0=4.64{\times}10^{-7}$ is fixed. Right: Line patterns
  for $\phi_0= 0.305$, $0.32$, $0.41$, $0.475$ and $0.499$ (from the
  top) indicated by dotted lines in the upper left panel.}
\mylab{f:vertcut}
\end{figure}
Next, we analyse regular line patterns as obtained from long-time
simulations \cite{FAT10_numdet}.  Excluding the initial transient we
take a sequence of $N$ regular deposition periods (lines), where $10
\lesssim N \lesssim 100$, depending on the period of the deposit and
required CPU time. We measure the amplitude, relative width (defined
as $2\sigma/[\mbox{period}]$, where $\sigma$ is the standard
deviation) and skewness of the lines, and the period of the line
pattern. We find that these quantities strongly depend on both the
evaporation number $\varOmega_0$ and the concentration $\phi_0$. We
focus on two cuts through region (d) in Fig.~\ref{f:phasediag} (dashed
lines).

First, we fix $\varOmega_0=4.64\times 10^{-7}$ and vary the bulk
concentration $\phi_0$.  Fig.~\ref{f:vertcut} presents line
characteristics and selected profiles. On increasing $\phi_0$ from a
region without periodic line deposition, one first finds large
amplitude almost solitary peaks separated by very large distances.  On
further increasing $\phi_0$, the amplitude first hardly changes and
later decreases. The period continuously decreases while the relative
width increases.  For higher $\phi_0$ the deposit pattern becomes
almost uniform, with a small amplitude harmonic modulation. Finally,
the amplitude goes to zero (at finite period) at the upper border of
the region (d).

\begin{figure}[t]
\includegraphics[width=0.477\hsize]{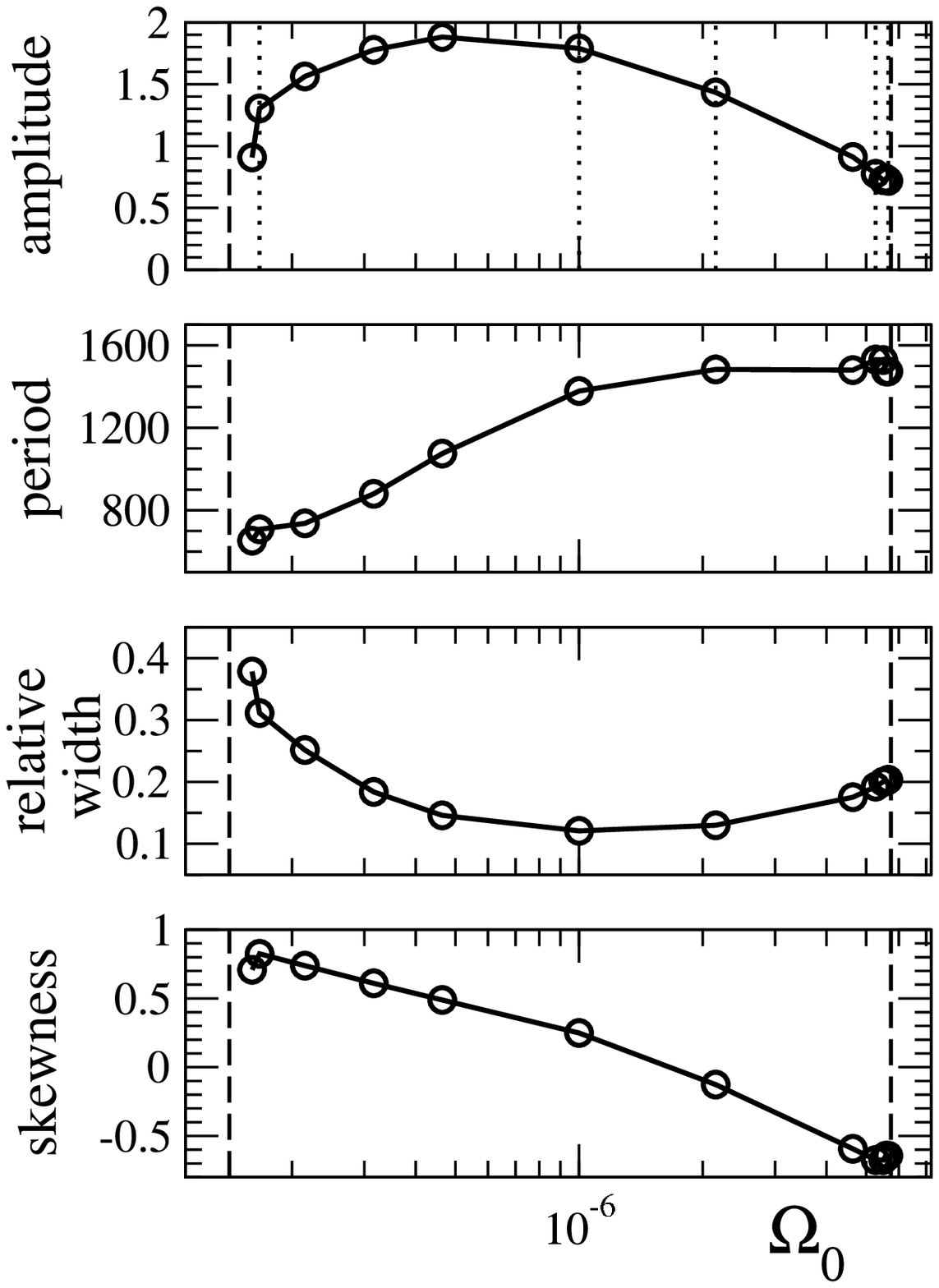}\hfill
\includegraphics[width=0.493\hsize]{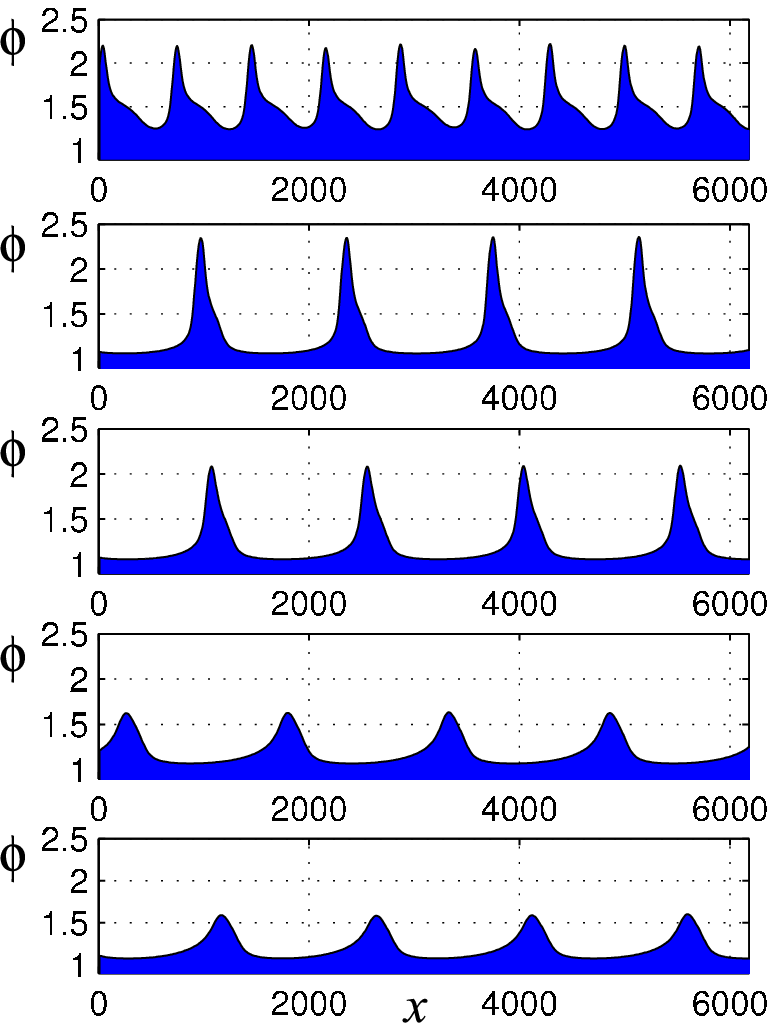}
\caption{{(color online) Left: dependence of the regular line pattern
    properties on $\varOmega_0$ in the range bounded by the vertical dashed lines
  (region (d) of Fig.~\ref{f:phasediag}).  $\phi_0=0.41$ is fixed. 
Right: Line patterns for $\varOmega_0= (0.17, 1.0, 2.15,
    5.27, 5.66)\times 10^{-6}$ (from the top) indicated by
    dotted lines in the upper left panel.}}
\mylab{f:horizcut}
\end{figure}

Second, Fig.\ \ref{f:horizcut} presents results for fixed
$\phi_0=0.41$ and varying $\varOmega_0$.  Increasing $\varOmega_0$,
moving from the narrow region (b) of multiple lines
(Fig.~\ref{f:phasediag}), one passes through a very narrow band of
intermittent line patterns (c) followed by the region (d) of regular
line patterns.  For the lowest values of $\varOmega_0$ in (d), the
patterns have a relatively small period and a small but non-zero
amplitude. The strongly anharmonic peaks are skewed to the right with
their tail pointing towards the receding front. On increasing
$\varOmega_0$ the period increases. The amplitude, however, first
increases and then decreases, until at a certain threshold the pattern
ceases to be periodic and we arrive in the narrow border region (b),
where only a finite number of lines are deposited.
Correspondingly, the relative width decreases, as lines get more
peaked.  Remarkably, the skewness changes sign, i.e.\ the tail of the
lines shifts from pointing towards the receding front, to pointing
away. This effect was observed in experiments on nanoparticle
suspensions \cite{philnote} and can be explained as follows: For
smaller values of $\varOmega_0$, the capillary ridge is large and
accommodates a large amount of nanoparticles as it recedes. When the
front pins, the capillary ridge is evaporated and these nanoparticles
are deposited in the thick tail to the right. When the liquid front
depins, it results in a further drop in the deposition thickness (seen
as shoulder in the tail). For higher values of $\varOmega_0$ the
capillary ridge is smaller and so the right tail is smaller.

%
%
In conclusion, we have studied a generic model for the
close-to-equilibrium deposition dynamics onto a surface from a polymer
solution or nanoparticle suspension. The model incorporates
wettability, capillarity, evaporation, convective transport of the
solution and diffusion of the solute and has been derived employing a
long-wave approximation. We find that a strong nonlinear dependence of
viscosity (i.e., the front mobility) on concentration triggers, in an
intricate interaction with evaporation and diffusion, the deposition
of periodic and aperiodic line patterns as observed in experiments for
many different materials and settings
\cite{Deeg00,HXL06,HXL07,YaSh05,Xu06,RDLL06,XXL07,BDG10,philnote}.
We believe that the model explains a basic mechanism for the formation
of regular line patterns.  They result from a self-organised cycle of
deposition-caused pinning-depinning events that is experimentally
often described as a stick-slip motion \cite{HXL06,Xu06,BDG10}.  In
the future the basic dynamical model should be extended to study the
influence of other important effects on the deposition.  These include
thermal effects, solute-dependent wettability and the nature of the
solvent-solute interaction.

We acknowledge support by the EU via the ITN MULTIFLOW
(PITN-GA-2008-214919).


\begin{thebibliography}{30}
\bibitem[{FAT()}]{FAT10_numdet}
\bibinfo{note}{The simulations start from the initial condition
  $h(x,0)=h_1+(h_2-h_1)H(x-x_0)$, where $h_1$ is the precursor film thickness,
  $h_2$ an upper stable film thickness \cite{LGP02} and $x_0\ll L$ the inital
  front position. $H(x)$ is the Heaviside step function and $\phi(x,0)=\phi_0$.
  Boundary conditions at $x=L$ are derived from asymptotics for
  $x\rightarrow\infty$. Details will be published elsewhere. Although the
  length $L$ of the computational domain is large, it is often too small for
  transients to decay. To follow the dewetting front we shift the computational
  domain many times by a fraction of $L$. To obtain the regular line patterns
  in Figs.~\ref{f:vertcut} and \ref{f:horizcut}, tens of shifts are necessary.
  Careful tests show that, in general, our results do not depend on the details
  of our numerical procedure.}
\end{thebibliography}
%
\end{document}